\documentclass[11pt]{article}
\usepackage{epsfig}

\begin{document}
\sffamily

\thispagestyle{empty}
\vspace*{15mm}
\begin{center}

{\LARGE \bf Complete spectra of the Dirac operator 

\vspace{2mm}
and their relation to confinement}

\vskip12mm
Falk Bruckmann$^a$, Christof Gattringer$^b$ and Christian Hagen$^a$

\vskip12mm
$^a$ Institut f\"ur Physik, Universit\"at Regensburg \\
93040 Regensburg, Germany 
\vskip3mm
$^b$ Institut f\"ur Physik, Unversit\"at Graz \\
8010 Graz, Austria 
\end{center}
\vskip30mm

\begin{abstract}
We compute complete spectra of the staggered lattice Dirac operator 
for quenched SU(3) gauge configurations below and above the critical 
temperature. The confined and the deconfined phase are characterized 
by a different response of the Dirac eigenvalues to a change of the 
fermionic boundary conditions. We analyze the role of the eigenvalues
in recently developed spectral sums representing the Polyakov loop. 
We show that the Polyakov loop gets its main contributions from the UV end 
of the spectrum. 
\end{abstract}

\vskip15mm
\begin{center}
{\sl To appear in Physics Letters B.}
\end{center}
\setcounter{page}0
\newpage
\noindent
{\large \bf Introduction}
\vskip2mm

In recent years many numerical studies of low lying eigenvalues and
eigenvectors of lattice Dirac operators were published. The reason for
analyzing low lying eigenvalues is twofold: Firstly, 
based on the Banks-Casher formula \cite{baca}, a close 
connection of topological objects to the chiral condensate has been 
proposed \cite{topochir}. Secondly, for low lying eigenvalues many
results from random matrix theory are available and were tested 
numerically in detail. 

While the low lying eigenvalues
and their relation to chiral symmetry breaking are
well analyzed, very little is known about the UV part of the spectrum and its
possible connection to confinement. Partly this situation is due the  
fact that an evaluation of complete spectra of a lattice Dirac 
operator is a considerable numerical challenge. 

A new way to analyze a possible relation between Dirac eigenvalues and
confinement was proposed in \cite{polspec}. It was shown that the Polyakov 
loop can be written as a linear combination of spectral sums over moments
of Dirac eigenvalues computed with different (fermionic) 
boundary conditions. For the
quenched case the Polyakov loop $P$ is an order parameter for confinement,
with $\langle P \rangle = 0$ in the confined phase, while in the 
deconfined phase (above $T_c$) the Polyakov loop develops a non-vanishing 
expectation value\footnote{In the theory with dynamical fermions 
one could study correlators of Polyakov loops to analyze the static 
potential.}. In \cite{polspec} it was speculated, that the difference between
confined and deconfined phase can be characterized by 
a different response of the 
Dirac eigenvalues to changing boundary conditions. This difference leads 
to a non-vanishing $\langle P \rangle$ only in the deconfined phase.

In this letter we analyze whether this scenario can be established in a 
numerical study of complete spectra of the staggered lattice Dirac operator. 
Furthermore we address the question which part of the spectrum, IR or UV, 
contributes most to the Polyakov loop.

\vskip5mm
\noindent
{\large \bf Spectral sums for the Polyakov loop}
\vskip2mm

The spectral sums for the Polyakov loop, first presented in \cite{polspec},
were derived for the Wilson Dirac operator. Here we use 
the staggered Dirac operator at vanishing quark mass,
\begin{equation}
D(n,m) \; = \; \frac{1}{2}
\sum_{\mu=1}^4 \, \eta_\mu(n) \, \Big[ 
U_\mu(n) \, \delta_{n+\hat{\mu},m} \, - \, U_\mu(n-\hat{\mu})^\dagger \,
\delta_{n-\hat{\mu},m} \, \Big] \; ,
\label{staggdir}
\end{equation}
where $n$ and $m$ are integer valued 4-vectors labeling the lattice sites
and $\eta_\mu(n) = (-1)^{n_1 + \, ... \, + n_{\mu-1}}$ is the staggered sign
function. The $U_\mu(n)$ denote the SU(3)-valued gauge links and we have set the 
lattice spacing to $a=1$. 

We use an $L^3 \times N$ lattice and require that the number $N$ of lattice
points in time direction (the 4-direction) is even. As it stands, 
the Dirac operator has periodic boundary conditions 
in all directions. Below we will
also need temporal boundary conditions with a phase $z$ or its complex
conjugate phase $z^\star$. These are implemented by multiplying the last
temporal links with $z$ and $z^\star$, i.e.,
$U_4(\vec{n},n_4\!=\!N) \rightarrow z \, U_4(\vec{n},n_4\!=\!N)$ or
$U_4(\vec{n},n_4\!=\!N) \, \rightarrow \, z^\star \, U_4(\vec{n},n_4\!=\!N)$.
Here we make the particular choice of $Z_3$-valued boundary conditions and
set $z = e^{i 2\pi/3}$.
We consider the Polyakov loop averaged over all of space,
\begin{equation}
P \; = \; \frac{1}{L^3} \sum_{\vec{n}} \, \mbox{Tr}_c \Big[
\prod_{n_4=1}^N  U_4(\vec{n},n_4) \Big] \; ,
\end{equation}
where $\mbox{Tr}_c$ denotes the trace over the color indices. 

Following the arguments in \cite{polspec}, one finds that the Polyakov loop is
given by a linear combination of spectral sums,
\begin{equation}
P \; =\; \frac{2^N}{3 N L^3} \, \bigg[
\sum_i \Big(\lambda^{(i)}\Big)^N \; + \; 
z^\star \sum_i \Big(\lambda^{(i)}_{z}\Big)^N \; + \; 
 z \sum_i \Big(\lambda^{(i)}_{z^\star}\Big)^N \bigg] \; .
\label{specsum}
\end{equation}
Each sum runs over all $3 L^3 N$ eigenvalues and $\lambda^{(i)},
\lambda^{(i)}_z$, and $\lambda^{(i)}_{z^\star}$ denote the eigenvalues 
computed with periodic, $z$-valued, and $z^\star$-valued boundary conditions
respectively. The Polyakov loop $P$ thus is represented as a linear
combination of spectral sums for the $N$-th power of the eigenvalues computed
with three different fermionic boundary conditions in time direction. The 
boundary conditions for the gauge fields are always kept periodic. 

It is interesting to note, that exact zero modes do not contribute to the
spectral sum for the Polyakov loop. Thus, isolated topological objects 
which give rise to a zero mode, do not play a 
role for building up the 
expectation value of the Polyakov loop\footnote{We remark that the staggered
Dirac operator does not have exact zero modes, but for sufficiently smooth
configurations the would-be zero modes can be identified relatively clearly
and are very close to the origin \cite{staggeredzm}.}. 

We stress that the result (\ref{specsum}) is an exact formula for the 
Polyakov loop of an arbitrary gauge configuration. Below we will
study numerically the expectation value $\langle P \rangle$ of the Polyakov
loop in a quenched ensemble. The r.h.s.\ of (\ref{specsum}) 
then is rewritten in terms 
of expectation values of the $N$-th moments of Dirac eigenvalues computed with
the three different boundary conditions,
\begin{equation}
\langle P \rangle  \; =\; \frac{2^N}{3 N L^3} \, \sum_i \bigg[
 \Big\langle \Big(\lambda^{(i)}\Big)^N \Big\rangle \; + \; 
z^\star \, \Big\langle \Big(\lambda^{(i)}_{z}\Big)^N \Big\rangle \; + \; 
 z \, \Big\langle \Big(\lambda^{(i)}_{z^\star}\Big)^N \Big\rangle 
\bigg] \; .
\label{specsumvev}
\end{equation}  
This formula relates the vacuum expectation value of the Polyakov loop, 
which originally is a purely gluonic quantity, to spectral sums of the Dirac
eigenvalues. Since the Polyakov loop is an order parameter for confinement
(in the quenched case), with $\langle P \rangle = 0$ in the confined phase 
($T < T_c$) and $\langle P \rangle \neq 0$ in the deconfined phase 
($T > T_c$), the formula (\ref{specsumvev}) allows to study the 
relation of confinement and the spectrum of the Dirac operator. 

As one crosses the critical temperature into the deconfined phase, 
$\langle P \rangle$ acquires a non-vanishing expectation value. 
Equation (\ref{specsumvev}) implies that the response of the
eigenvalues to the boundary conditions has to change at $T_c$, such that the
spectral sums on the right-hand side do no longer cancel and 
$\langle P \rangle \neq 0$. In the subsequent sections we study the
response of the eigenvalues to changing boundary conditions and explore
which parts of the Dirac
spectrum give the main contributions to the spectral sums.

\vskip5mm
\noindent
{\large \bf Distribution of the Dirac eigenvalues 
and their response to changing boundary conditions}
\vskip2mm

For a numerical study of the formula (\ref{specsumvev}) we need to compute
complete spectra of the Dirac operator using three different
boundary conditions. Since a numerical evaluation of all eigenvalues is 
a demanding task we are restricted to relatively small lattices. Here we 
use quenched gauge field configurations generated with the
L\"uscher-Weisz action \cite{LuWeact} on lattices of size $6^3 \times 4$.

We work with two different values of the inverse coupling, 
$\beta = 7.6$ and $\beta = 8.0$. Setting the scale with the Sommer 
parameter one finds lattice spacings of $a = 0.194(4)\,$fm and   
$a = 0.135(1)\,$fm for the two couplings \cite{scale}. With a temporal 
extension of $N=4$ this corresponds to temperatures of $T = 254$ MeV
for $\beta = 7.6$, and $T = 364$ MeV for  $\beta = 8.0$. Thus, we have two 
ensembles with temperatures below and above the QCD phase transition which for
the quenched case is at $T_c \sim 300$ MeV. Using LAPACK routines
we compute complete spectra of the staggered Dirac operator with the three
boundary conditions for 2000 configurations from each of the two ensembles. 
The statistical errors we quote for the averaged observables
are evaluated with single elimination Jackknife.

\begin{figure}[t]
\hspace*{-4mm}
\includegraphics[height=5.7cm,clip]{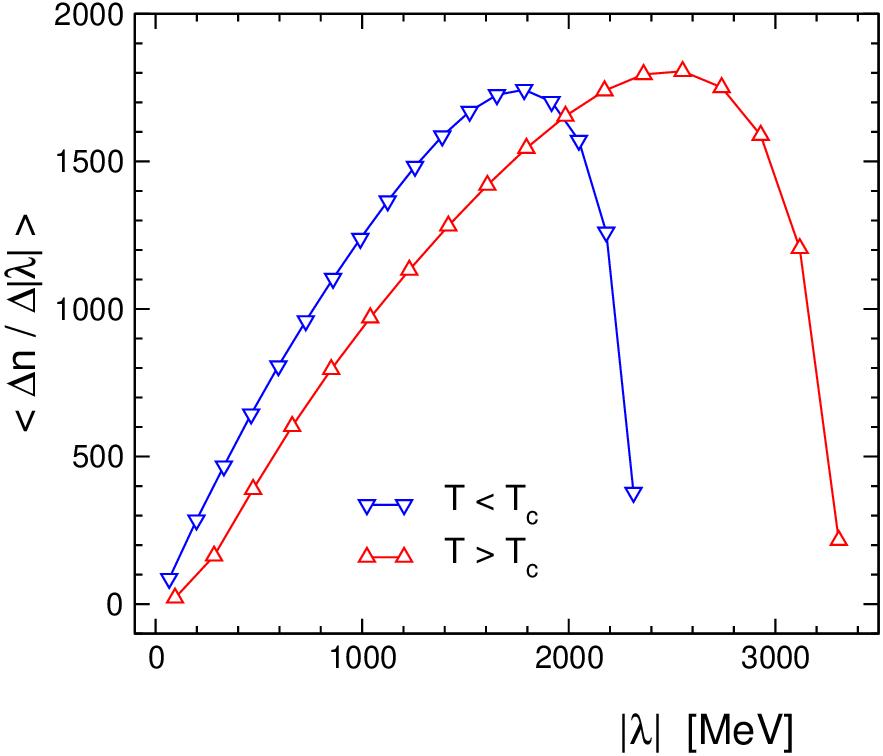}
\includegraphics[height=5.7cm,clip]{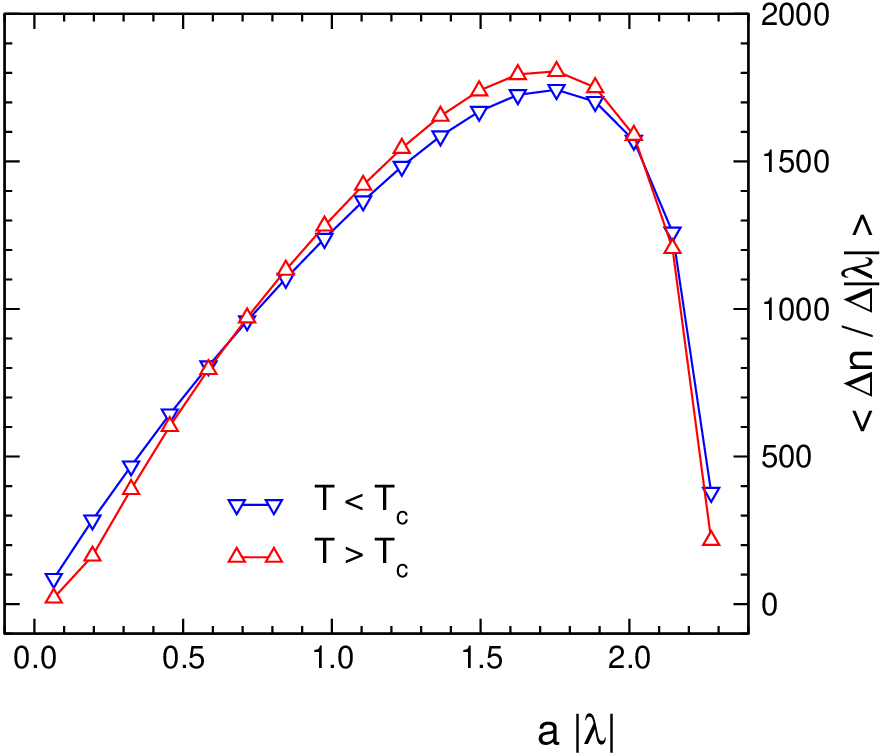}
\caption{
Distribution of the eigenvalues $\lambda$ as a function of
$|\lambda|$ (for periodic b.c.). In the l.h.s.\ plot we use MeV as unit for 
the eigenvalues, while on the r.h.s.\ the dimensionless quantity
$a |\lambda|$ (lattice units) is plotted on the horizontal axis.  
}
\label{fig_distribution}
\end{figure} 

We begin the presentation of our numerical results with a discussion of the
eigenvalue distribution. Since the massless staggered Dirac operator is an
anti-hermitian matrix, it has eigenvalues on the imaginary axis. We order 
these with respect to their absolute value and the sign of the 
imaginary part (the eigenvalues come in
complex conjugate pairs). For analyzing the distribution 
of the eigenvalues we divide $|\lambda|$ into small bins of size 
$\Delta |\lambda|$ and count the number of eigenvalues, $\Delta n$, in each of
the bins. In Fig.~\ref{fig_distribution} we plot $\Delta n/\Delta |\lambda|$
as a function of $|\lambda|$ for both ensembles below and above $T_c$ 
(see \cite{tilo} for a similar analysis in the case of SU(2)). 
On the horizontal axis we use either physical units (l.h.s.\ plot), 
or the dimensionless combination $a |\lambda|$ (r.h.s.). 

As can be seen from the r.h.s.\ plot, the curves for the distribution of
the eigenvalues are very similar for the two ensembles when plotted in
lattice units. At small $|\lambda|$ the density of eigenvalues is a little
bit depleted for the ensemble with $T > T_c$. This can be understood 
qualitatively from the Banks-Casher formula \cite{baca}, which predicts a 
vanishing spectral density at the origin for the chirally symmetric phase 
(above $T_c$). The corresponding 
opening of a spectral gap is well documented in
numerical studies \cite{bacagap} and this phenomenon is reflected in the 
lower density at small $|\lambda|$ seen in our data for $T > T_c$. 
The area under the two curves has to be equal (= the total number of
eigenvalues) and we observe a light enhancement of the density for the 
$T > T_c$ spectra near the maximum. When we use physical units on the
horizontal scale (l.h.s.\ plot), 
the two densities are stretched with different factors due
to the different lattice spacing for the two values of the coupling we use.
   
It is obvious that the distribution of the eigenvalues plays an important role 
for the spectral representation (\ref{specsumvev}). Energy ranges where the
density of eigenvalues is large will in general be more important than 
those parts of the spectrum with a low density. In order to disentangle the role of the 
density from other aspects, such as the response to changing boundary
conditions, we now study two observables for individual eigenvalues.

\begin{figure}[t]
\hspace*{-3mm}
\includegraphics[height=5.7cm,clip]{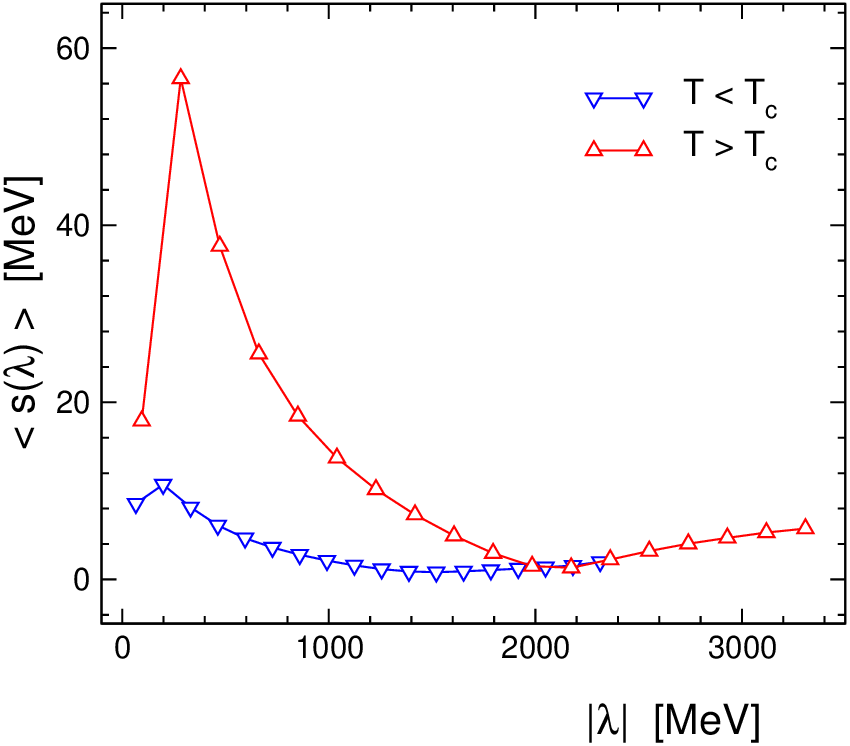}
\includegraphics[height=5.7cm,clip]{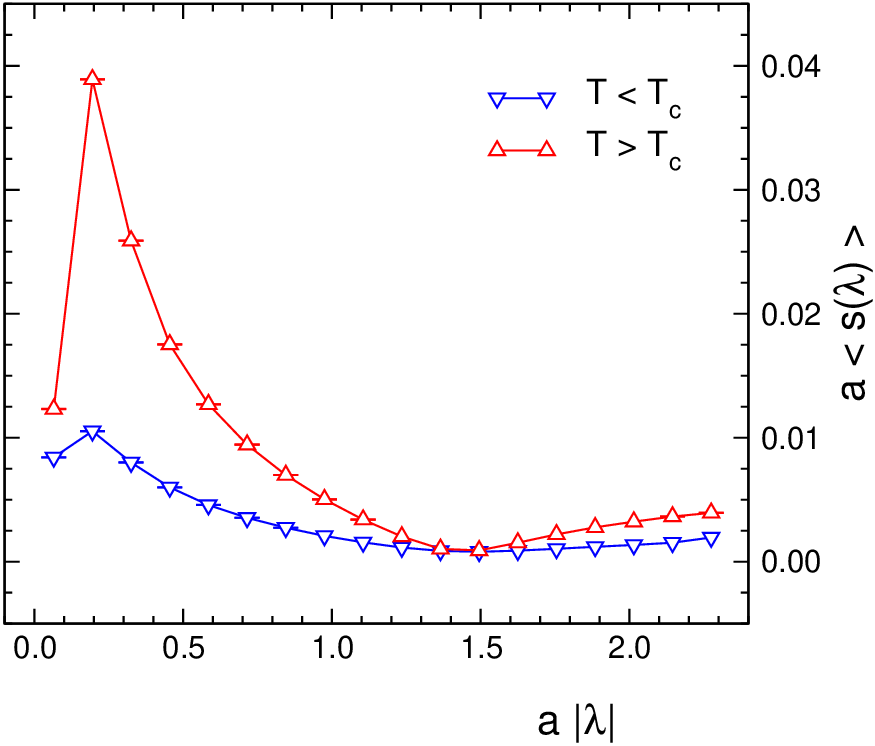}
\caption{
Average shift of the eigenvalues when changing the boundary condition
of the Dirac operator, plotted as a function of 
$|\lambda|$. In the l.h.s.\ plot physical units are used
and lattice units on the r.h.s.
}
\label{fig_shift}
\end{figure} 

The first property of the eigenvalues we consider is their average shift when
the boundary conditions are changed. To quantify this effect we define the
average shift $s(\lambda^{(i)})$, by comparing an  
eigenvalue $\lambda^{(i)}$ for periodic boundary conditions 
to its partners $\lambda^{(i)}_z$ and $\lambda^{(i)}_{z^\star}$, computed
with $z$- and $z^\star$-boundary conditions,
\begin{equation}
s(\lambda^{(i)}) \; = \; \Big( \,
|\lambda^{(i)} - \lambda^{(i)}_z| \, + \,
|\lambda^{(i)} - \lambda^{(i)}_{z^\star}| \, + \,
|\lambda^{(i)}_z - \lambda^{(i)}_{z^\star}| \, \Big) \, / \, 3 \; .
\end{equation}
In Fig.\ \ref{fig_shift} we show the average shift $s(\lambda)$
as a function of $|\lambda|$, again using physical units on the 
l.h.s.\ plot and lattice units on the r.h.s. The most obvious feature of the
plots is the fact that the shift of the eigenvalues is
considerably stronger for the $T > T_c$ ensemble, in particular towards the IR
end of the spectrum. This confirms an observation made in \cite{gapresponse},
where it was shown, that for $T > T_c$ the size of the spectral gap strongly
depends on the fermionic boundary condition. The plots furthermore
show, that for $T>T_c$ 
the shift is stronger than for the data at $T< T_c$, not only in the deep IR,
but for all of the eigenvalues. 

Apart from the different total shift, the two ensembles display also common
features: A clear maximum close to the IR end, a minimum for midrange values
and another increase at the UV end. The plots demonstrate that the IR modes
are shifted most, combined with a less pronounced shift of the largest
eigenvalues. It is interesting to note that the parts of the spectrum with 
large shifts coincide with low densities (compare Fig.\
\ref{fig_distribution}). 

Finally, we remark that both curves show a drop for the first 
bin in the IR. We attribute this to the would-be zero modes which are 
stabilized by topology and thus should not move at all. However, since the
staggered Dirac operator has only approximate zero modes, these eigenvalues
move a little bit, but considerably less than the bulk modes, thus creating
the drop in the lowest bin. 

\vskip5mm
\noindent
{\large \bf Contributions to the Polyakov loop}
\vskip2mm

We have demonstrated that when changing the boundary conditions,
different parts of the spectrum are shifted by
different amounts. However, in the formula (\ref{specsumvev})
for the Polyakov loop the $N$-th powers of the eigenvalues enter and 
the shifted spectra are
weighted with the phases $z$ and $z^\star$. Thus, we now consider the
contribution $c(\lambda^{(i)})$ of an 
individual eigenvalue to the spectral sum,
\begin{equation}
c(\lambda^{(i)}) \; = \; 
\frac{2^N}{3 N L^3} \, \bigg[
\Big(\lambda^{(i)}\Big)^N \; + \; 
z^\star \, \Big(\lambda^{(i)}_{z}\Big)^N \; + \; 
z \,\Big(\lambda^{(i)}_{z^\star}\Big)^N \, \bigg] \; .
\end{equation} 
  
\begin{figure}[t]
\hspace*{-6mm}
\includegraphics[height=5.7cm,clip]{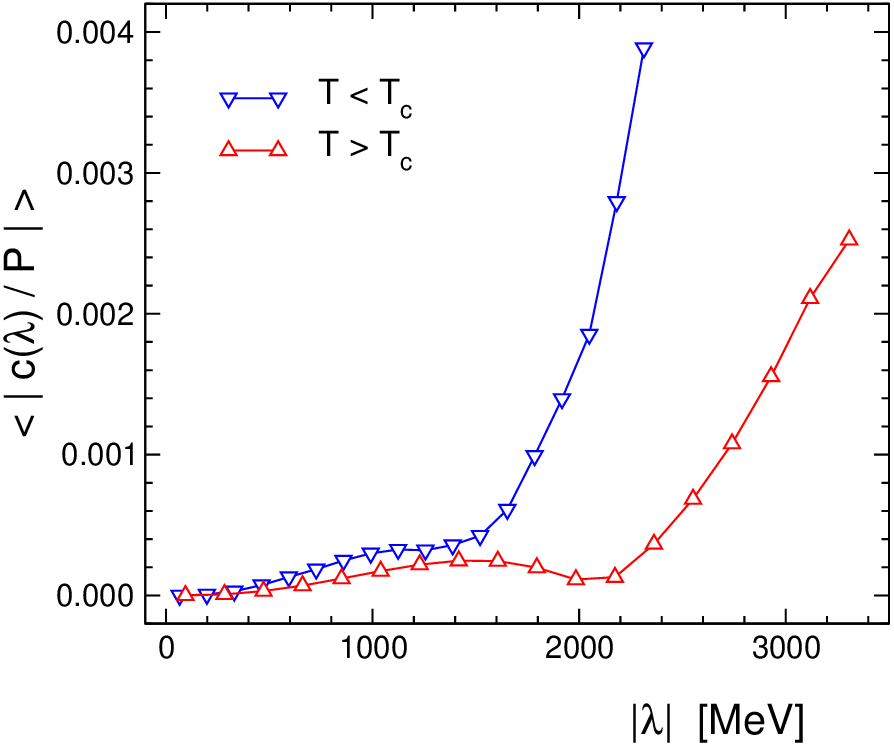}
\includegraphics[height=5.7cm,clip]{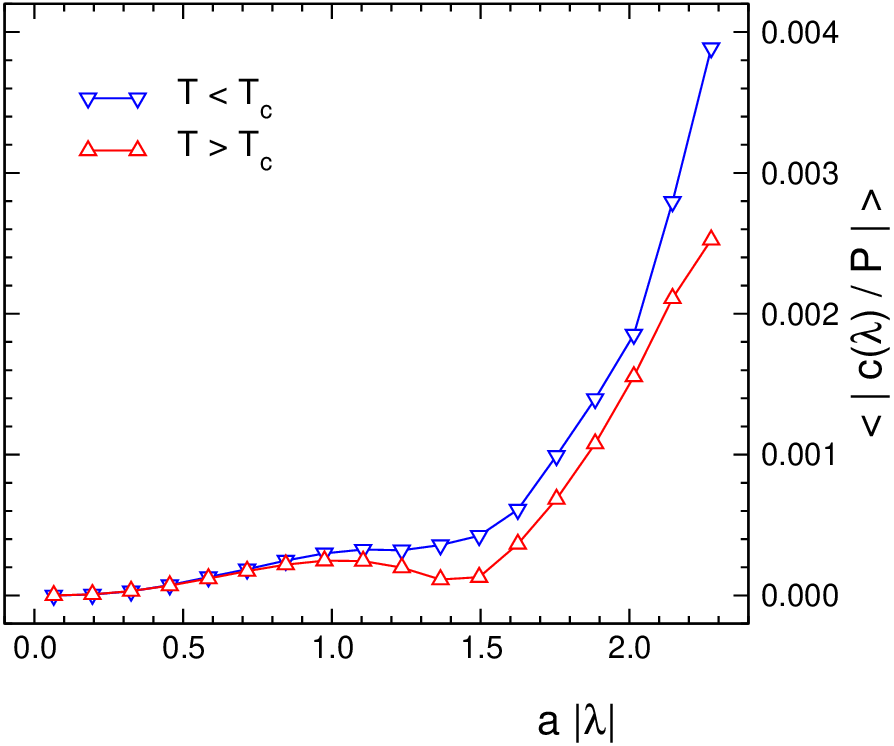}
\caption{
Contribution of an eigenvalue to the spectral sum for the Polyakov loop,
as a function of $|\lambda|$
(l.h.s.\ plot: physical units, r.h.s.\ plot: lattice units). 
}
\label{fig_contrib}
\end{figure} 
In Fig.\ \ref{fig_contrib} we show the absolute value of the
contribution $c(\lambda)$ normalized by the total Polyakov 
loop\footnote{We remark, that on a finite lattice the Polyakov
loop does not vanish exactly also below $T_c$. However, this ``microscopic
value'' vanishes in the thermodynamic limit.} as a function 
of $|\lambda|$. For both ensembles the size of the contribution increases 
strongly towards the UV end of the spectrum. Furthermore, when plotted in
lattice units (r.h.s.\ plot), the two curves show a similar
behavior. They both start out with a modest slope which, after a small dip (or shoulder)
near $a |\lambda| \sim 1.5$, turns into a steeper ascent. 
Fig.\ \ref{fig_contrib} demonstrates, that although according to the last 
plot the IR modes are shifted more, the UV modes have a larger contribution 
to the Polyakov loop.

\begin{figure}[t]
\hspace*{-3mm}
\includegraphics[height=5.7cm,clip]{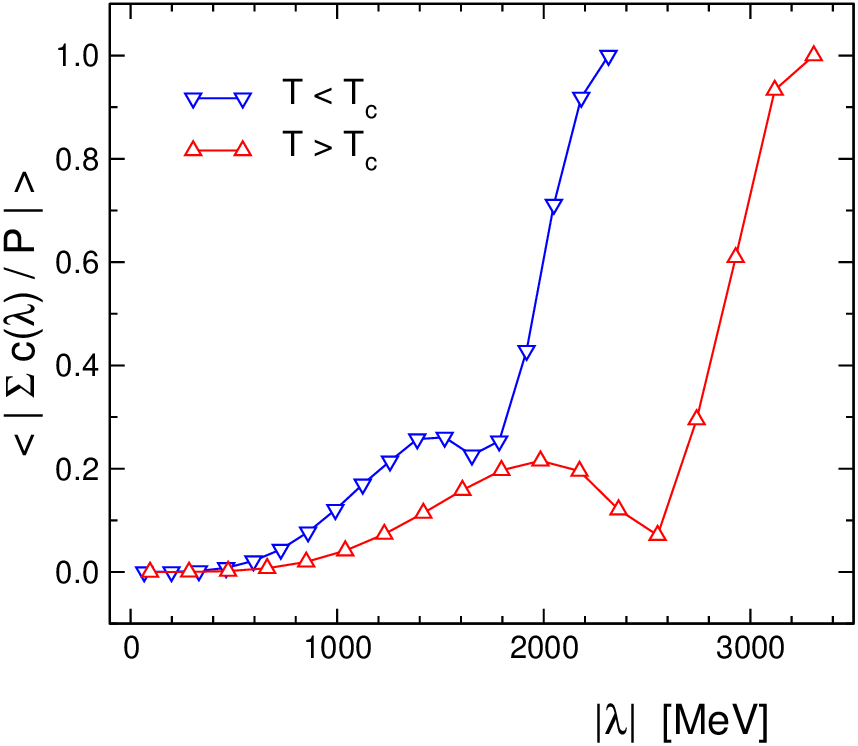}
\includegraphics[height=5.7cm,clip]{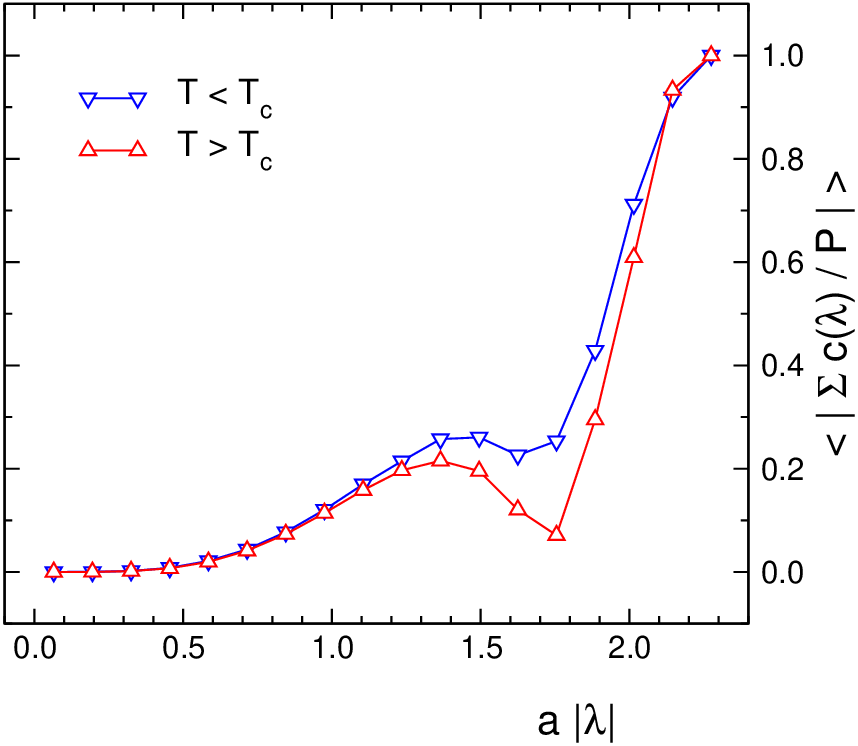}
\caption{
Accumulated contribution to the spectral sum for the Polyakov loop,
as a function of $|\lambda|$
(l.h.s.\ plot: physical units, r.h.s.\ plot: lattice units).}
\label{fig_accum}
\end{figure}

Having studied the contribution of the individual modes, we finally have to
fold in also the distribution of the eigenvalues analyzed in 
Fig.\ \ref{fig_distribution}. To take the spectral density 
into account we compute
the accumulation of the contributions $c(\lambda)$, by summing all 
contributions 
up to a given value of $|\lambda|$, i.e., we analyze the cumulated quantity
$\sum_{|\lambda^\prime| \leq |\lambda|} c(\lambda^\prime)$. 
In Fig.\ \ref{fig_accum} we plot the absolute value of this quantity,   
normalized with the total Polyakov loop, as a function 
of $|\lambda|$. As was to be expected, the IR modes do not contribute a lot
and most of the Polyakov loop is carried by the UV modes. What is somewhat
surprising, is the fact that the cumulation is not a monotonically increasing
function, but has a dip which is particularly pronounced for $T > T_c$. This
shows that the buildup of the Polyakov loop from the spectral sums is not a
simple linear process. The r.h.s.\ plot demonstrates that, when plotted in 
lattice units, the curves for the two ensembles are rather similar, as could
already be concluded from the similarity of the behavior of the corresponding
curves in Figs.\ \ref{fig_distribution} and  \ref{fig_contrib}. We
remark that at the UV end both curves reach 1.0 to machine precision, which is
a good check that the exact formulas (\ref{specsum}), (\ref{specsumvev}) were
implemented correctly.

\vskip5mm
\noindent
{\large \bf Phase of the Polyakov loop above $T_c$}
\vskip2mm

\begin{figure}[t]
\hspace*{-3mm}
\includegraphics[height=3.8cm,clip]{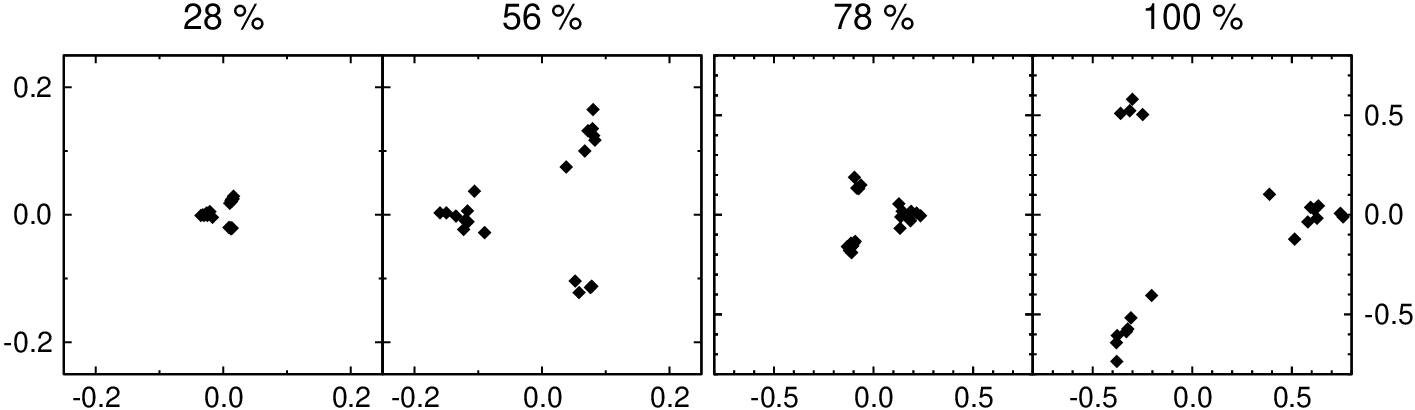}
\caption{
The Polyakov loop in the complex plane for 20 configurations above $T_c$ 
as reconstructed from the spectral sums
with 28\%, 56\%, 78\% and 100\% of the eigenvalues (left to right). Note that
the two plots on the l.h.s.\ have a different scale.}
\label{complexpol}
\end{figure}

Above $T_c$ the phase of the Polyakov loop has values close to the angles
of the group center, i.e., close to $0\,, \;2\pi/3$ and $4\pi/3$. This can be 
seen in the plot on the very right in Fig.~\ref{complexpol} where we show
the complete spectral sum for the Polyakov loop for 20 gauge configurations. 
For the truncated sum,
we observe a surprising phenomenon: When including less than
roughly two thirds of the eigenvalues, one finds that the results for the 
Polyakov loop show a phase shift of 180 degrees (see the two plots on the l.h.s.\
of Fig.~\ref{complexpol}). Only the largest third of the eigenvalues 
has the correct phase and, since the UV contributions dominate, determine the 
final phase (the two plots on the r.h.s.\ of Fig.~\ref{complexpol}). 
 
\begin{figure}[b!]
\hspace*{-1mm}
\includegraphics[height=5.6cm,clip]{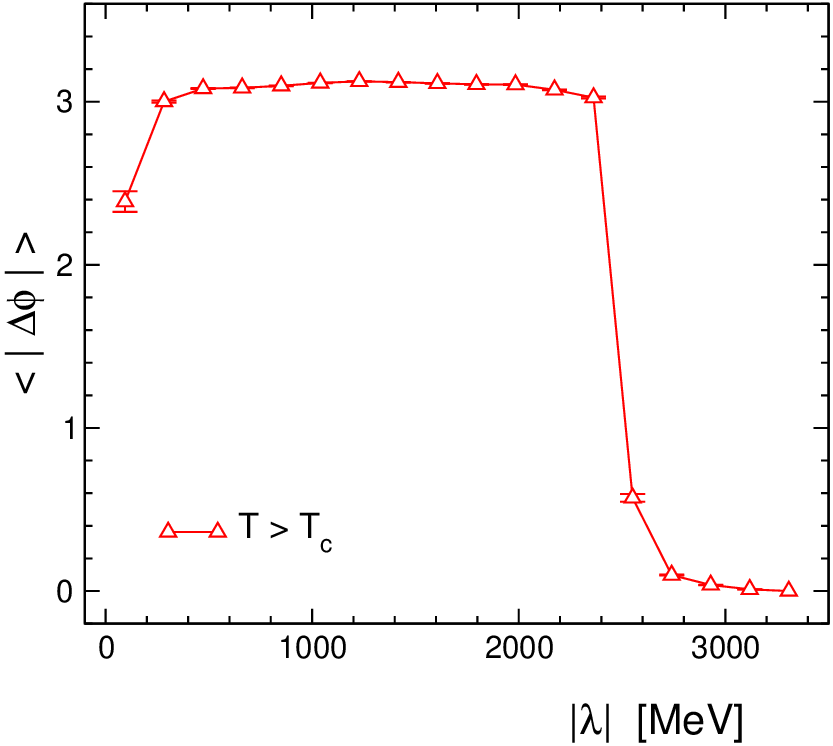}
\includegraphics[height=5.6cm,clip]{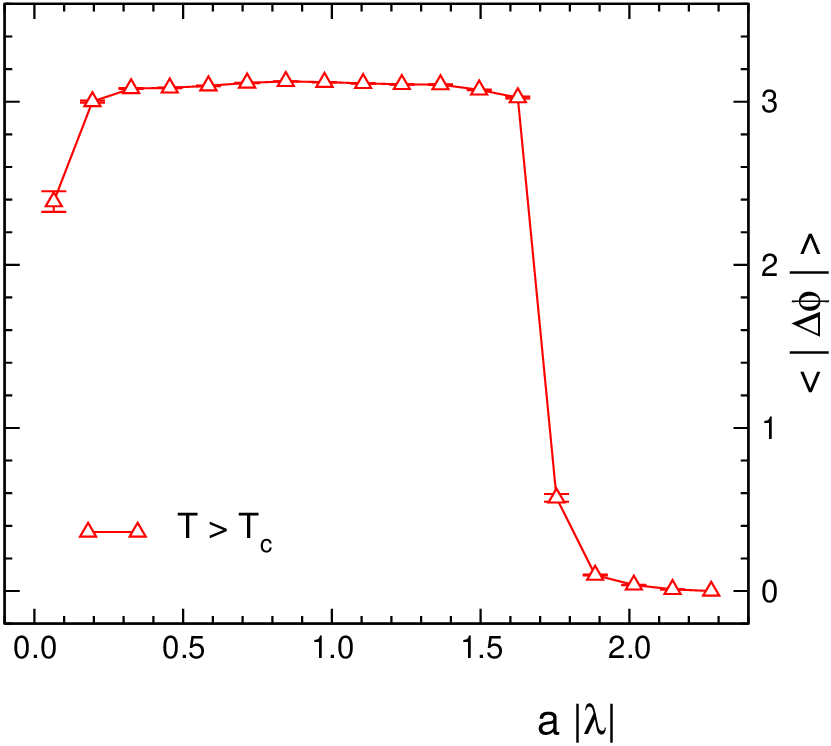}
\caption{Phase shift of the truncated spectral sums for the Polyakov loop.}
\label{polphase}
\end{figure} 

In Fig.~\ref{polphase} we show the 
phase shift $\Delta \phi$ of the truncated spectral sum relative to the  
phase of the full sum. It is obvious, that when truncating the
sum at less than two thirds of the eigenvalues, the phase is shifted by 
a value of 180 degrees (i.e., a shift of $\pi$), and only the UV eigenvalues drive
the relative shift to zero. It is interesting to note, that the position where
the phase starts to come out right coincides with the position of the
dip observed in Fig.~\ref{fig_accum}. To summarize,
the truncated sum for every configuation
first evolves in the opposite direction in the complex 
plane and after two thirds turns back to approach the correct full Polyakov loop.

\vskip5mm
\noindent
{\large \bf Summary and interpretation}
\vskip2mm

We have generalized the discussion of spectral sums of Dirac eigenvalues 
representing the Polyakov loop \cite{polspec} to the case of the staggered 
Dirac operator. In order to study these spectral sums numerically we computed complete
Dirac spectra with three different fermionic boundary conditions,
using quenched ensembles below and above the QCD phase transition. 

Different aspects of these spectra were studied, 
in particular the distribution of the eigenvalues and their shift under a
change of boundary conditions were analyzed. Concerning this shift 
we established that the IR modes are shifted most and also towards the UV end
found a small increase. A comparison with the results from the distribution
analysis showed that the eigenvalues with largest shift coincide with the 
regions of lowest eigenvalue density. Qualitatively this pattern holds for 
both ensembles, but for $T > T_c$ the shift of the eigenvalues is 
considerably larger than for $T< T_c$. This enhanced 
response to changing boundary conditions leads to a non-vanishing
$\langle P \rangle$ for $T < T_c$, while in the confined phase only a
microscopic value of $\langle P \rangle$ emerges which vanishes in the
thermodynamic limit.
  
Concerning the buildup of the Polyakov loop, 
we considered the contribution of an individual
eigenvalue as well as the accumulated contribution. The contribution of
individual eigenvalues does not take into account that the density of the 
eigenvalues is a function of their size and thus disentangles the two effects
of a varying density and the different contribution of individual eigenvalues. 
Both the individual as well as the accumulated contributions show that  
mainly the eigenvalues in the UV build up the Polyakov loop. For the phase 
of the accumulated contribution we have shown that in the IR 
a phase shift of 180 degrees is observed which vanishes at the UV end.  

We have established the following qualitative scenario for the
relation between Dirac eigenvalues and the Polyakov loop below and above
$T_c$: In both phases the eigenvalues respond to changing the boundary
conditions, but the response is considerably larger in the deconfined
phase. As a consequence, the linear combination of the spectral 
sums with different boundary conditions leads to $\langle P \rangle > 0$ 
in that case. Concerning the role of different eigenvalues, we find
that the spectral sums for $\langle P \rangle$ are dominated 
by the UV end of the spectrum. 

\newpage
\noindent
{\bf Acknowledgments:}
We thank Stefan Keppeler, Christian Lang, Andreas Sch\"afer, 
Erhard Seiler, Kim Splittorff, Jac Verbaarschot, Tilo Wettig
and Andreas Wipf
for interesting discussions. This work is supported by BMBF.

\end{document}